\documentstyle[12pt]{article}

\newcommand{\be}{\begin{equation}}
\newcommand{\ee}{\end{equation}}

\newcommand{\dlt}{\delta}
\newcommand{\prt}{\partial}
\newcommand{\br}{{\bf r}}
\newcommand{\bk}{{\bf k}}
\newcommand{\bt}{\beta}

\newcommand{\ep}{\varepsilon}
\newcommand{\al}{\alpha}
\newcommand{\ra}{\rightarrow}
\newcommand{\sgm}{\sigma}

\newcommand{\om}{\omega}

\newcommand{\dgr}{\dagger}
\newcommand{\lbd}{\lambda}
\newcommand{\Lbd}{\Lambda}
\newcommand{\cF}{{\cal F}}

\begin{document}

\begin{center}

{\Large {\bf Self-Consistent Theory of Bose-Condensed Systems} \\ [5mm]

V.I. Yukalov} \\ [3mm]

{\it Bogolubov Laboratory of Theoretical Physics, \\
Joint Institute for Nuclear Research, Dubna 141980, Russia \\ 
and \\
Theoretische Physik, Universit\"at Essen, \\
Universit\"atsstrasse 5, Essen 45117, Germany}

\end{center}

\begin{abstract}

In the theory of Bose-condensed systems, there exists the well known problem, 
the Hohenberg-Martin dilemma of conserving versus gapless approximations. This 
dilemma is analysed and it is shown that it arises because of the internal 
inconsistency of the standard grand ensemble, as applied to Bose-systems with 
broken global gauge symmetry. A solution of the problem is proposed, based on 
the notion of representative statistical ensembles, taking into account all 
constraints imposed on the system. A general approach for constructing 
representative ensembles is formulated. Applying a representative ensemble 
to Bose-condensed systems results in a completely self-consistent theory, both 
conserving and gapless in any approximation.

\end{abstract}

\vskip 1cm


\section{Introduction}

Bose-Einstein condensation in Bose systems is commonly associated with 
the global gauge symmetry breaking. The most general way of the gauge 
symmetry breaking is by means of the Bogolubov shift in the field 
operators [1--4]. In his first works [1,2], Bogolubov considered a 
weakly-nonideal Bose gas in the limit of a small condensate depletion, 
which can occur at low temperatures and asymptotically weak interactions. 
The following generalization of the Bogolubov ideas, based on the 
standard grand ensemble with broken gauge symmetry, confronted the well 
known problem of theories being either nonconserving or exhibiting an 
unphysical gap in the spectrum. Hohenberg and Martin [5] were, probably, 
the first who analysed in detail this dilemma and classified the approaches 
onto conserving versus gapless. In a nonconserving theory, the dynamic 
conservation laws are not valid and general thermodynamic relations do 
not hold, which is related to the thermodynamic instability of the system. 
If the theory is conserving, but exhibits a gap in the spectrum, it also 
has incorrect thermodynamics and contradicts the theorems of the gapless 
spectrum by Bogolubov [4] and Hugenholtz and Pines [6]. A thorough 
discussion of the Hohenberg-Martin dilemma has recently been done by 
Andersen [7], where the extensive literature on the problem is cited.

It is important to emphasize that the inconsistency, happening in the 
theory, is internal, but not connected with the employed approximations. 
Really, suppose one applies the standard grand ensemble, with a chemical 
potential $\mu$, to an equilibrium Bose-condensed system. After breaking 
the gauge symmetry, by means of the Bogolubov shift, one has to minimize 
the grand thermodynamic potential with respect to the number of condensed 
particles $N_0$. Bogolubov [1--4] illustrated this procedure for a weakly 
nonideal gas. Ginibre [8] proved that this procedure is general for 
arbitrary Bose systems. The Bogolubov-Ginibre minimization procedure 
defines the chemical potential $\mu\ra\mu_{BG}(\rho,T)$ as a function 
of density $\rho$ and temperature $T$.

On the other hand, there exists the Hugenholtz-Pines relation [6] 
also defining the chemical potential $\mu\ra\mu_{HP}(\rho,T)$. The 
most general proof on this relation, for any Bose system and arbitrary 
temperatures, is based on the consideration of increments of statistical 
averages under gauge variations [4]. This is analogous to the consideration 
of the Ward-Takahashi identity. The Hugenholtz-Pines relation follows 
from the symmetry of the Hamiltonian, but is not connected with the system 
stability.

Thus, there are two representations for the chemical potential. 
One, $\mu_{BG}$, is a consequence of the condition of thermodynamic 
stability, while another, $\mu_{HP}$, results from the symmetry properties. 
Two different, both physically and mathematically, conditions do not need to 
define one and the same quantity, so that generally, $\mu_{BG}\neq\mu_{HP}$.
And this is exactly what happens [5,7], making the theory not self-consistent.

So, it is not some particular approximations that are guilty for the arising 
inconsistency. Also, the gauge symmetry breaking is not to be blamed. The 
concept of symmetry breaking under phase transitions is rather general [9,10]. 
But the inconsistency, related to the inequality $\mu_{BG}\neq\mu_{HP}$, is 
generic for the used statistical ensemble.

The Hohenberg-Martin dilemma has been addressed in a number of papers. 
One usually attempts to cure the problem in a mean-field approximation 
complemented by some phenomenological tricks. The most common, and the 
least justified, trick is by omitting the anomalous averages. One often 
ascribes this trick to Popov, terming it the "Popov approximation". First 
of all, however, it is not difficult to infer from the original works by 
Popov [11--13], cited in this regard, that he has never suggested or used 
such an unjustified trick. The person who really first proposed and employed 
the trick was Shohno [14]. This was known in literature as Shohno model. 
But in the case of the broken gauge symmetry the anomalous averages become 
principally important and cannot be omitted. Direct calculations [15] show 
that the anomalous averages at low temperatures can be even much larger than 
the normal ones. So, neglecting the terms that are larger than what is left 
cannot be a reasonable mathematical approximation. In addition, the omission 
of anomalous averages renders the system thermodynamically unstable [15,16], 
with incorrect thermodynamics [17].

To remove the gap in the spectrum, one sometimes resorts to what Bogolubov 
[4] called the mismatch of approximations. Then one defines the chemical 
potential $\mu_{BG}$ from the condition of thermodynamic stability in 
one approximation, but, for cancelling the appearing gap, one invokes 
the self-energies in another, higher, approximation, or just adding 
phenomenological terms. As is evident, the mismatch of approximations 
is not a logical procedure, is not uniquely defined, and does not solve 
the problem that for each given approximation the theory remains not 
self-consistent. Also, the mismatch of approximations leads to incorrect 
thermodynamics, with a first-order phase transition, instead of the 
second-order one [18--20]. Thermodynamic self-consistency can be preserved 
only when all quantities are derived from the same Hamiltonian in the same 
chosen approximation [21,22].

As is mentioned above, the inconsistency, manifesting itself in the 
inequality $\mu_{BG}\neq\mu_{HP}$, is generic for the used statistical 
ensemble and does not depend on the employed approximations. The problem 
can be solved by using a representative statistical ensemble, which takes 
into account all constraints imposed on the Bose-condensed system [23]. 
Then, for instance, the Hartree-Fock-Bogolubov (HFB) approximation can 
be made conserving and gapless [24].

In the present paper, the general notion of representative statistical 
ensembles is formulated for arbitrary statistical systems. The representative 
ensemble for a Bose-condensed system is defined. It is shown that the theory, 
based on the representative ensemble, is completely self-consistent, 
independently of the used approximations. This solves the Hohenberg-Martin 
dilemma, suggesting an approach that is free from inconsistencies and 
paradoxes.

Everywhere, below the system of units is used, where $\hbar\equiv 1$ and 
$k_B\equiv 1$.

\section{Representative Ensembles}

It was already Gibbs [25] himself who mentioned that for a correct description 
of a statistical system, in addition to a canonical distribution, one has to 
accurately take into account all conditions and constraints, imposed on the 
system and uniquely representing its properties. From this general idea stems 
the term "representative ensembles", which was used by ter Haar [26,27]. For 
equilibrium systems, one defines representative ensembles through the 
conditional maximization of the Gibbs entropy, as was done by Janes [28]. 
A generalization for quasi-equilibrium systems was given in Ref. [29]. Here, 
we describe a general method for constructing representative ensembles for 
both equilibrium and nonequilibrium systems.

An {\it equilibrium statistical ensemble} is a pair $\{\cF,\hat\rho\}$ 
of a space of microstates $\cF$ and a statistical operator $\hat\rho$ on this 
space. Let a set $\{\hat C_i\}$ of self-adjoint operators $\hat C_i$ be given, 
which we shall call the condition operators, since these define the {\it 
statistical conditions}
\be
\label{1}
C_i =\; <\hat C_i>\; = \; {\rm Tr}\hat\rho \; \hat C_i \; ,
\ee
which are assumed to be imposed on the considered system. The trace 
operation in Eq. (1) is over the space of microstates $\cF$. One obvious 
constraint is the normalization condition for the statistical operator,
\be
\label{2}
1 \; = \; < \hat 1_\cF> \; = \; {\rm Tr}\hat\rho \; ,
\ee
where $\hat 1_\cF$ is the unity operator in $\cF$. Another standard 
condition is the definition of the internal energy
\be
\label{3}
E \; = \; < \hat H> \; = \; {\rm Tr}\hat\rho \; \hat H
\ee
as the average of a Hamiltonian $\hat H$. However, in addition to these 
standard conditions, other statistical conditions (1) can also be necessary 
for correctly representing the system.

For the conditional maximization of the Gibbs entropy 
$S\equiv-{\rm Tr}\hat\rho\ln\hat\rho$, under statistical conditions (1), (2), 
and (3), it is convenient to introduce [30] the {\it information functional}
\be
\label{4}
I[\hat\rho] ={\rm Tr}\hat\rho\ln\hat\rho + \lbd_0 \left ( 
{\rm Tr}\hat\rho - 1\right ) +\bt\left ( {\rm Tr}\hat\rho \hat H - 
E \right ) + \bt \sum_i \nu_i \left ( {\rm Tr}\hat\rho \hat C_i -
C_i \right ) \; ,
\ee
with the Lagrange multipliers $\lbd_0$, $\bt$, and $\bt\nu_i$. The statistical 
operator is given by the minimum of the information functional (4),
\be
\label{5}
\hat\rho = \frac{1}{Z} \; e^{-\bt H} \; ,
\ee
with the {\it grand Hamiltonian}
\be
\label{6}
H \equiv \hat H + \sum_i \nu_i \hat C_i \; .
\ee
Here $Z\equiv\exp(1+\lbd_0)$ is the partition funtion and $\bt\equiv T^{-1}$ 
is inverse temperature.

The {\it representative statistical ensemble} for an equilibrium system 
under constraints (1), (2), and (3), is the pair $\{\cF,\hat\rho\}$ of a 
space of microstates $\cF$ and the statistical operator (5) with the grand 
Hamiltonian (6).

Note that the condition operators $\hat C_i$ are to be self-adjoint, but 
they do not need to be compulsorily the integrals of motion. Thus, the 
number-of-particle operator $\hat N$ is not an integral of motion, when 
the gauge symmetry is broken.

A {\it nonequilibrium statistical ensemble} is a triplet 
$\{\cF,\hat\rho,\prt/\prt t\}$ of a space of microstates $\cF$, the 
statistical operator $\hat\rho\equiv\hat\rho(0)$ at the initial time $t=0$, 
and of a prescribed evolution law, symbolically denoted here as $\prt/\prt t$. 
The law of evolution can be given as the Liouville equation for the 
time-dependent statistical operator $\hat\rho(t)$ or as the Heisenberg 
equations of motion for physical operators, or in some other form. A general 
way of deriving the evolution equations is by extremizing an effective action 
functional with respect to the given dynamical variables [31].

To construct an action functional, one defines the corresponding Lagrangian
\be
\label{7}
\hat L \equiv \hat E - \hat H \; ,
\ee
in which $\hat E$ is the time-dependent energy operator, containing the 
time derivative. The effective action, for a system with the Lagrangian (7), 
under constraints (1), is
\be
\label{8}
A = \int \left ( \hat L - \nu_i \hat C_i \right ) dt \; ,
\ee
where $\nu_i$ are the Lagrange multipliers. Equation (8) can be rewritten as
\be
\label{9}
A = \int \left ( \hat E - H\right ) dt \; ,
\ee
with the same grand Hamiltonian (6). The evolution equations follow from 
the extremization of the effective action, $\dlt A=0$. For instance, if the 
effective action $A=A[\psi]$ is a functional of field operators $\psi(\br,t)$ 
and $\psi^\dgr(\br,t)$, then the latter satisfy the equations
\be
\label{10}
\frac{\dlt A[\psi]}{\dlt\psi^\dgr(\br,t)} = 0 \; , \qquad
\frac{\dlt A[\psi]}{\dlt\psi(\br,t)} = 0 \; .
\ee
The grand Hamiltonian (6) governs both the statistical properties of 
equilibrium systems and the temporal evolution of nonequilibrium systems.

\section{Bose Systems}

Let us consider a one-component system of spinless particles characterized 
by the field operators $\psi$ and $\psi^\dgr$ satisfying the Bose commutation
relations. For the system above the critical temperature $T_c$, the space 
of microstates is the Fock space $\cF(\psi)$ generated by the field operators 
[30--32]. In addition to statistical conditions (2) and (3), there is one 
normalization condition for the total number of particles $N=<\hat N>$. Hence, 
the grand Hamiltonian (6) is $H=\hat H-\mu\hat N$, where $\mu$ is the usual 
chemical potential.

However, the situation changes below the critical temperature $T_c$, when the 
global gauge symmetry becomes broken. The symmetry breaking is realized by the 
Bogolubov shift
\be
\label{11}
\psi(\br,t) \; \longrightarrow \; \hat\psi(\br,t) \; \equiv \; 
\eta(\br,t) +\psi_1(\br,t) \; ,
\ee
in which $\eta(\br,t)$ is the condensate wave function and $\psi_1(\br,t)$ 
is the field operator of uncondensed particles. The field operators $\psi_1$ 
and $\psi_1^\dgr$ generate the Fock space $\cF(\psi_1)$, which is the space 
of microstates for the system with broken gauge symmetry. The spaces of 
microstates $\cF(\psi)$, for the system above $T_c$, and $\cF(\psi_1)$, for 
the system below $T_c$, are mutually orthogonal [33,34].

For the system with broken gauge symmetry, there are now two variables, 
the condensate wave function $\eta$ and the field operator of uncondensed 
particles $\psi_1$. These variables are assumed to be linearly independent, 
being orthogonal to each other,
\be
\label{12}
\int \eta^*(\br,t) \psi_1(\br,t)\; d\br = 0 \; .
\ee
Respectively, instead of one normalization condition for the total number 
of particles, we have now two normalization conditions, for the number of 
condensed particles
\be
\label{13}
N_0 \; = \; <\hat N_0> \; ,
\ee
where, the average is over the space $\cF(\psi_1)$, 
$\hat N_0\equiv N_0\hat 1_\cF$, $\hat 1_\cF$ being a unity operator in 
$\cF(\psi_1)$, and
\be
\label{14}
N_0  = \int |\eta(\br,t)|^2 d\br \; ,
\ee
and for the number of uncondensed particles
\be
\label{15}
N_1 \; = \; <\hat N_1> \; ,
\ee
with the operator of uncondensed particles
\be
\label{16}
\hat N_1 \equiv \int \psi_1^\dgr(\br,t)\psi_1(\br,t)\; d\br \; .
\ee
The total number of particles is $N=N_0+N_1$.

With the broken gauge symmetry, the average $<\psi_1>$ might be nonzero. 
This, however, would mean that quantum numbers, such as spin or momentum, 
would not be conserved. To avoid this, one implies the constraint
\be
\label{17}
<\psi_1(\br,t)>\; = \; 0 \; ,
\ee
guaranteeing the conservation of quantum numbers. Constraint (17) can 
be represented in the standard form (1) of a statistical condition by 
introducing the self-adjoint operator
\be
\label{18}
\hat\Lbd \equiv \int \left [ \lbd(\br,t)\psi_1^\dgr(\br,t) + 
\lbd^*(\br,t)\psi_1(\br,t) \right ] \; d\br \; ,
\ee
in which $\lbd(\br,t)$ is a complex function. Then constraint (17) is 
equivalent to the {\it quantum conservation condition}
\be
\label{19}
<\hat \Lbd > \; = \; 0 \; .
\ee
Thus, for a system with broken gauge symmetry, there are three 
statistical conditions, (13), (15) and (19). To construct the 
corresponding representative ensemble, we have to follow the general 
prescription of Sec. 2.

For a system in equilibrium, in view of Eq. (4), the information 
functional is
$$
I[\hat\rho] = {\rm Tr}\hat\rho \ln\hat\rho + 
\lbd_0 \left ( {\rm Tr}\hat\rho - 1\right ) + 
\bt \left ( {\rm Tr}\hat\rho \hat H -  E\right ) -
$$
\be
\label{20}
 - \bt\mu_0 \left ( {\rm Tr}\hat\rho \hat N_0 - N_0 \right ) - 
\bt\mu_1 \left ( {\rm Tr}\hat\rho \hat N_1 - N_1 \right ) - 
\bt {\rm Tr}\hat\rho \hat \Lbd \; .
\ee
The minimization of this functional yields the statistical operator 
(5) with the grand Hamiltonian
\be
\label{21}
H = \hat H - \mu_0 \hat N_0 - \mu_1 \hat N_1 - \hat\Lbd \; ,
\ee
in agreement with Eq. (6).

In the general case of a nonequilibrium system, we need to consider 
the Lagrangian (7), in which the temporal energy operator is
$$
\hat E = \int \left [ \eta^*(\br,t)\; i\; \frac{\prt}{\prt t}\;
\eta(\br,t) + \psi_1^\dgr(\br,t) \; i \; \frac{\prt}{\prt t} \;
\psi_1(\br,t) \right ] \; d\br \; .
$$
The action functional (9) has to be extremized with respect to two 
linearly independent variables, $\eta$ and $\psi_1$, which implies
\be
\label{22}
\frac{\dlt A[\eta,\psi_1]}{\dlt\eta^*(\br,t)} \; = \; 0 \; ,
\qquad \frac{\dlt A[\eta,\psi_1]}{\dlt\psi_1^\dgr(\br,t)} \; = 
\; 0 \; .
\ee
The latter is equivalent to the equations
\be
\label{23}
i\; \frac{\prt}{\prt t}\; \eta(\br,t) = 
\frac{\dlt H[\eta,\psi_1]}{\dlt\eta^*(\br,t)} \; , \qquad
i\; \frac{\prt}{\prt t}\; \psi_1(\br,t) = 
\frac{\dlt H[\eta,\psi_1]}{\dlt\psi_1^\dgr(\br,t)} \; .
\ee

The average of the grand Hamiltonian (21) is
\be
\label{24}
< H > \; = \; E -\mu_0 N_0 - \mu_1 N_1 \; .
\ee
If in experiment only the total number of particles, or the total average 
density, can be fixed, then average (24) should be expressed in the 
standard form $<H>=E-\mu N$ through the system chemical potential 
$\mu$. Comparing this with Eq. (24) defines the chemical potential 
\be
\label{25}
\mu = \mu_0 n_0 + \mu_1 n_1 
\ee
through the Lagrange multipliers $\mu_0$ and $\mu_1$ and through the 
fractions $n_0\equiv N_0/N$ and $n_1\equiv N_1/N$.

Let us take the Hamiltonian energy operator in the usual form
$$
\hat H = \int \hat\psi^\dgr(\br) \left ( -\; \frac{\nabla^2}{2m} +
U \right ) \hat\psi(\br) \; d\br \; +
$$
\be
\label{26}
+ \; \frac{1}{2} \; \int \hat\psi^\dgr(\br) \hat\psi^\dgr(\br') 
\Phi(\br-\br') \hat\psi(\br') \hat\psi(\br)\; d\br d\br' \; ,
\ee
in which $\hat\psi(\br)=\hat\psi(\br,t)$ is the Bogolubov shifted field 
operator $\hat\psi\equiv\eta+\psi_1$, defined in Eq. (11). Here $U=U(\br,t)$ 
is an external potential. The interaction potential $\Phi(\br)$ is symmetric, 
so that $\Phi(-\br)=\Phi(\br)$, and is assumed to possess the Fourier 
transform. With the Bogolubov shift (11), Hamiltonian (26) can be represented 
as a sum of five terms, $H=H^{(0)}+H^{(1)}+H^{(2)}+H^{(3)}+H^{(4)}$, depending 
on the powers of the operators $\psi_1$ and $\psi_1^\dgr$ in each of them [34]. 
The zero-order term $H^{(0)}$ contains no $\psi_1$. The necessary and sufficient 
condition for the validity of constraint (19) is the absence in the Hamiltonian 
of terms linear in $\psi_1$ [35]. This requires to define the Lagrange 
multipliers $\lbd(\br,t)$ such that $H^{(1)}=0$.

The evolution equations (23) give
$$
i\; \frac{\prt}{\prt t}\; \eta(\br,t) = \left ( -\; \frac{\nabla^2}{2m} +
U - \mu_0 \right ) \eta(\br,t) + \int \Phi(\br-\br') 
\left [ |\eta(\br')|^2 \eta(\br) + \hat X(\br,\br')\right ] \; d\br \; ,
$$
$$
i\; \frac{\prt}{\prt t}\; \psi_1(\br,t) = \left ( -\; \frac{\nabla^2}{2m} +
U - \mu_1 \right ) \psi_1(\br,t) + 
$$
\be
\label{27}
+\int \Phi(\br-\br') 
\left [ |\eta(\br')|^2 \psi_1(\br) + \eta^*(\br')\eta(\br)\psi_1(\br') +
\eta(\br')\eta(\br)\psi_1^\dgr(\br') +
\hat X(\br,\br')\right ] \; d\br' \; ,
\ee
where, for brevity, the time dependence in the integrands is not shown 
explicitly and the notation for the correlation operator
$$
\hat X(\br,\br') \equiv \psi_1^\dgr(\br') \psi_1(\br')\eta(\br) +
$$
$$
+ \psi_1^\dgr(\br') \eta(\br') \psi_1(\br) + 
\eta^*(\br') \psi_1(\br') \psi_1(\br) +
\psi_1^\dgr(\br') \psi_1(\br') \psi_1(\br)
$$
is used. It is straightforward to check that the evolution equations (27) 
guarantee the validity of all local conservation laws on the operator level, 
as it should be for the equations derived from a variational procedure.

Averaging the first of Eqs. (27) over $\cF(\psi_1)$, we obtain the explicit 
equation for the condensate wave function
$$
i\; \frac{\prt}{\prt t}\; \eta(\br,t) = \left ( -\; \frac{\nabla^2}{2m} +
U - \mu_0 \right ) \eta(\br,t) + 
$$
\be
\label{28}
+\int \Phi(\br-\br')  \left [ \rho(\br') \eta(\br) + 
\rho_1(\br,\br') \eta(\br') + \sgm_1(\br,\br')\eta^*(\br') + 
\xi(\br,\br')\right ] \; d\br' \; ,
\ee
in which
$$
\rho_1(\br,\br') \equiv \; <\psi_1^\dgr(\br') \psi_1(\br) > \; , \qquad
\sgm_1(\br,\br') \equiv \; <\psi_1(\br') \psi_1(\br) > \; , 
$$
$$
\xi(\br,\br') \equiv \; <\psi_1^\dgr(\br') \psi_1(\br')\psi_1(\br) > \; ,
$$
$$
\rho(\br) = \rho_0(\br) +\rho_1(\br) \; , \qquad 
\rho_0(\br) \equiv |\eta(\br)|^2 \; , \qquad
\rho_1(\br) \equiv\rho_1(\br,\br) \; = \; < \psi_1^\dgr(\br)\psi_1(\br)> \; .
$$

Equation (28) is valid for an arbitrary Bose system. For a system in 
equilibrium, we require that the condensate function be time independent, 
$\prt\eta/\prt t=0$. If, in addition, the system is uniform, with no external 
fields, then Eq. (28) yields
\be
\label{29}
\mu_0 = \rho \Phi_0 + \int \Phi(\br) \left [ \rho_1(\br,0) +
\sgm_1(\br,0) + \frac{\xi(\br,0)}{\sqrt{\rho_0}}\right ] \; d\br \; .
\ee

The matrix Green function $G(12)$ for uncondensed particles is defined in the 
usual way [4--6], where the common abbreviation can be employed, denoting the 
set $\{\br_j,t_j\}$ by a single number $j$. The elements of this matrix are
$$
G_{11}(12) = - i<\hat T\psi_1(1)\psi_1^\dgr(2)> \; , \qquad
G_{12}(12) = - i<\hat T\psi_1(1)\psi_1(2)> \; ,
$$
\be
\label{30}
G_{21}(12) = - i<\hat T\psi_1^\dgr(1)\psi_1^\dgr(2)> \; , \qquad
G_{22}(12) = - i<\hat T\psi_1^\dgr(1)\psi_1(2)> \; ,
\ee
where $\hat T$ is the chronological operator. For an equilibrium system, 
we may pass to the Fourier transform $G(\bk,\om)$ of $G(12)$. Equations of 
motion give
$$
G_{11}(\bk,\om) = 
\frac{\om+k^2/2m+\Sigma_{11}(\bk,\om)-\mu_1}{D(\bk,\om)} \; , 
$$
\be
\label{31}
G_{12}(\bk,\om) = - \; \frac{\Sigma_{12}(\bk,\om)}{D(\bk,\om)} = 
G_{21}(\bk,\om) \; ,
\ee
where $\Sigma(\bk,\om)=\left [\Sigma_{\al\bt}(\bk,\om)\right ]$ is the 
self-energy matrix, and the denominator is given by the equations
$$
D(\bk,\om) \equiv \Sigma_{12}^2(\bk,\om) - 
G_{11}^{-1}(\bk,\om)G_{11}^{-1}(\bk,-\om) \; ,
$$
$$
G_{11}^{-1}(\bk,\om) = \om\;  -\; \frac{k^2}{2m} \; - \; 
\Sigma_{11}(\bk,\om) + \mu_1 \; .
$$
From the Bogolubov theorem [4], we have
\be
\label{32}
\left | G_{11}(\bk,0) - G_{12}(\bk,0)\right | \geq 
\frac{mn_0}{k^2} \; .
\ee
Using here Eqs. (31), we obtain the Hugenholtz-Pines relation
\be
\label{33}
\mu_1 =\Sigma_{11}(0,0) - \Sigma_{12}(0,0) \; ,
\ee
which makes the spectrun gapless.

Comparing Eqs. (29) and (33), we see that these expressions are rather 
different, and there is no reason to require that $\mu_0$ be equal to 
$\mu_1$. As an illustration, we can resort to the Hartree-Fock-Bogolubov 
(HFB) approximation, in which we find
\be
\label{34}
\mu_0 =\rho\Phi_0 + \frac{1}{V}\; 
\sum_{p\neq 0} (n_p+\sgm_p)\Phi_p \; , \qquad
\mu_1 =\rho\Phi_0 + \frac{1}{V}\; 
\sum_{p\neq 0} (n_p - \sgm_p)\Phi_p \; ,
\ee
where $V$ is the system volume, $\Phi_p$ is the Fourier transform of 
$\Phi(\br)$ and $n_p$ and $\sgm_p$ are the normal and anomalous averages, 
respectively, in the momentum space,
$$
n_k \; \equiv \; < a_k^\dgr a_k > \; , \qquad 
\sgm_k \; \equiv \; <a_k a_{-k}> \; ,
$$
with $a_k$ being the Fourier transform of $\psi_1(\br)$. Clearly, 
$\mu_0\neq\mu_1$. The anomalous average $\sgm_k$, as can be shown by 
direct calculations [15,24], can be much larger than the normal average 
$n_k$. Hence, there is no any sense in omitting $\sgm_k$. As has also 
been mentioned above, the presence of $\sgm_k$ is crucially important 
for the stability of the system [15,16,23,36].

The physics of dilute Bose gases has resently attracted a great deal 
of attention, both experimentally and theoretically (see review works 
[7,37--40]). These gases are described by the local interaction potential 
\be
\label{35}
\Phi(\br) = \Phi_0\dlt(\br) \; , \qquad \Phi_0 \equiv 4\pi\;
\frac{a_s}{m} \; ,
\ee
in which $a_s$ is the scattering length. For this potential, Eqs. (34) 
reduce to
\be
\label{36}
\mu_0 = (\rho +\rho_1 +\sgm_1)\Phi_0 \; , \qquad
\mu_1 =(\rho + \rho_1 - \sgm_1) \Phi_0 \; ,
\ee
where
$$
\rho_1 = \frac{1}{V} \; \sum_{k\neq 0} n_k \; , \qquad
\sgm_1 = \frac{1}{V} \; \sum_{k\neq 0} \sgm_k \; .
$$
The particle spectrum has the Bogolubov form
$$
\ep_k = \sqrt{(ck)^2 +\left ( \frac{k^2}{2m}\right )^2} \; ,
$$
with the sound velocity $c$ defined by the equation
\be
\label{37}
mc^2 = (\rho_0 +\sgm_1) \Phi_0 \; ,
\ee
where $\rho_0\equiv N_0/V$ is the condensate density. The anomalous average 
$\sgm_1$, invoking the dimensional regularization, can be represented as
\be
\label{38}
\sgm_1 = \frac{(mc)^2}{\pi^2}\; \sqrt{m\rho_0\Phi_0} \; - \;
\frac{(mc)^3}{2\sqrt{2}\pi} \; \int_0^\infty \;
\frac{(\sqrt{1+x^2}-1)^{1/2}}{\sqrt{1+x^2}} \left [ {\rm coth}
\left ( \frac{mc^2}{2T}\right ) -1 \right ] \; dx \; .
\ee
The condensate fraction becomes
\be
\label{39}
n_0 =  1 \; - \; \frac{(mc)^3}{3\pi^2\rho} \left\{ 1 + 
\frac{3}{2\sqrt{2}}\; \int_0^\infty \; \left ( 
\sqrt{1+x^2}-1\right )^{1/2} \left [{\rm coth}\left (
\frac{mc^2}{2T}\; x\right ) - 1 \right ]\; dx \right \} \; .
\ee
And for the superfluid fraction, defined through the response to a velocity 
boost [37], we obtain
\be
\label{40}
n_s =  1 \; - \; \frac{(mc)^3mc^2}{6\sqrt{2}\pi^2\rho T} \int_0^\infty \; 
\frac{(\sqrt{1+x^2}-1)^{3/2}x\; dx}{\sqrt{1+x^2}{\rm sinh}^2(mc^2x/2T)} \; .
\ee
The above equations show that the system undergoes the second-order phase 
transition at the critical temperature
$$
T_c = \frac{2\pi}{m} \left [
\frac{\rho}{\zeta(3/2)}\right ]^{2/3}
$$
coinciding with that for the ideal gas, as it should be for a mean-field 
approximation with the local interaction potential (35). Note that all 
integrals in Eqs. (38), (39), and (40) are well defined and convergent.

In conclusion, the old-standing problem, the Hohenberg-Martin dilemma, 
is analysed, which classifies the theories of Bose-condensed systems 
onto either conserving or gapless. This dilemma is shown to be of general 
nature, not dependent on the used approximations, and having its footing 
in the generic inconsistency of the standard grand ensemble applied to 
the Bose systems with broken global gauge symmetry. The notion of 
representative statistical ensembles for arbitrary systems is formulated. 
A representative ensemble uniquely represents the given statistical system 
by taking into account all constraints and conditions imposed on the system. 
The usage of representative ensembles for Bose-condensed system provides 
the remedy from internal inconsistencies, cures old paradoxes, and gives 
the solution to the Hohenberg-Martin dilemma. The suggested theory for 
Bose-condensed systems, based on a representative ensemble, is completely 
self-consistent, conserving, and gapless. This is achieved by introducing 
two Lagrange multipliers, $\mu_0$ and $\mu_1$, which are responsible for 
two normalization conditions, for $N_0$ and $N_1$, respectively. Note that 
in the current experiments with trapped atoms [37--40] two atomic numbers 
can be controlled, the total number of atoms $N$ and the number of condensed
atoms $N_0$. In addition, the operator $\hat N_0$ is the integral of motion, 
since $[\hat N_0,H]=0$. So, fixing two normalization conditions is justified 
both mathematically as well as physically. Two Lagrange multipliers are 
necessary in order to uniquely define the condensate wave function, which 
is a two-component order parameter. This agrees with the general rule for 
any statistical system: 

{\it To uniquely define an order parameter of a system, it is necessary to 
introduce as many Lagrange multipliers as is the number of components in the 
order parameter}.

\vskip 5mm

{\bf Acknowledgement}. I appreciate financial support from the German
Research Foundation. I am grateful for fruitful discussions to M. Girardeau,
R. Graham, H. Kleinert, and E.P. Yukalova.

\end{document}